%Paper: hep-th/9508156
%From: Fay Dowker <dowker@cosmic.physics.ucsb.edu>
%Date: Mon, 28 Aug 95 21:53:34 -0700

%-------------------------
% This paper uses harvmac
%-------------------------
\input harvmac
\noblackbox
\def\Title#1#2{\rightline{#1}\ifx\answ\bigans\nopagenumbers\pageno0\vskip1in
\else\pageno1\vskip.8in\fi \centerline{\titlefont #2}\vskip .5in}

scaled\magstep3
 
scaled\magstep3
%%%%%%%%%%%%%%%%%%%

%-------------------
%  definitions
%-------------------
%

\font\cmss=cmss10 \font\cmsss=cmss10 at 7pt
\def\hf{{1 \over 2}}

\def\tr{{\rm Tr}}
\def\ahat{{ \hat a}}
\def\bhat{{\hat b}}
\def\chat{{\hat c}}
\def\m{M_2}
\def\mtil{{{\tilde M}_2}}
\def\mz{M_2^0}
\def\mtilz{{{\tilde M}_2^0}}

\def\p{\partial}
\def\IR{\relax{\rm I\kern-.18em R}}
\def\IC{\relax\hbox{$\inbar\kern-.3em{\rm C}$}}
\def\semidirect{\mathbin{\hbox{\hskip2pt\vrule height 4.1pt depth -.3pt
width .25pt \hskip-2pt$\times$}}}
\def\inbar{\,\vrule height1.5ex width.4pt depth0pt}
\def\IZ{\relax\ifmmode\mathchoice
{\hbox{\cmss Z\kern-.4em Z}}{\hbox{\cmss Z\kern-.4em Z}}
{\lower.9pt\hbox{\cmsss Z\kern-.4em Z}}
{\lower1.2pt\hbox{\cmsss Z\kern-.4em Z}}\else{\cmss Z\kern-.4em
Z}\fi}

%
%-------------------
% references
%-------------------
%
\lref\callias{C. Callias, Comm. Math. Phys. {\bf 62} (1978) 213. }
\lref\dhvw{L. Dixon, J. A. Harvey, C. Vafa and E. Witten, Nucl. Phys.
{\bf B261} (1985)
678; Nucl. Phys. {\bf B274} (1986) 285. }
\lref\gradry{ I. S. Gradshteyn and I.M. Ryzhik,
``Tables of Integrals, Series and Products,''
(Academic Press, New York, 1965.) }
\lref\hitchin{N. Hitchin, Adv. Math. {\bf 41} (1974) 1.}
\lref\egh{T. Eguchi, P. B. Gilkey, A. J. Hanson, Phys. Rept. {\bf 66}
(1980) 213.}
\lref\ll{L. D. Landau and E. M. Lifshitz, Quantum Mechanics, Pergamon, (1977).}
\lref\fr{U. Fano and G. Racah, Irreducible Tensorial Sets, Academic
Press, (1959).}
\lref\ag{L. Alvarez-Gaume, J.Phys. {\bf A16} (1983) 4177.}
\lref\jr{R. Jackiw and C. Rebbi, Phys. Rev. {\bf D13} (1976) 3398.}
\lref\gaunt{J. P. Gauntlett, Nucl. Phys. {\bf B411} (1994) 443.}
\lref\gaunttwo{J. P. Gauntlett, Nucl. Phys. {\bf B400} (1993) 103.}
\lref\gm{G. W. Gibbons and N. Manton, Nucl. Phys. {\bf B274}
 (1986) 183.}
\lref\HL{J. A. Harvey and J. Liu, Phys. Lett. {\bf B268} (1991) 40.}
\lref\banks{T. Banks, M. Dine, H. Dijkstra and W. Fischler,
Phys. Lett. {\bf B212} (1988) 45.}
\lref\gibb{G. W. Gibbons and P. J. Ruback, Commun. Math. Phys. {\bf 115}
(1988) 267.}
\lref\ms{N. S. Manton and B. Schroers, Ann. Phys. {\bf 225} (1993) 290.}
\lref\manton{N. S. Manton, Phys. Lett. {\bf B110} (1982) 54.}
\lref\dab{A. Dabholkar and J. A. Harvey, Phys. Rev. Lett.  {\bf 63} (1989) 719
\semi
A. Dabholkar, G. Gibbons, J. A. Harvey, and F. Ruiz Ruiz,
Nucl. Phys. B {\bf B340} (1990) 33.}
\lref\sena{A. Sen, Int. J. Mod. Phys. {\bf A9} (1994) 3707.}
\lref\senb{A. Sen, Phys. Lett. {\bf 329} (1994) 217.}
\lref\GNO{P. Goddard, J. Nuyts and D. Olive, Nucl. Phys. {\bf B125}
(1977) 1.}
\lref\WO {E. Witten and D. Olive, Phys. Lett. {\bf 78B} (1978) 97. }
\lref\CSF{ E. Cremmer, J. Scherk and S.  Ferrara,
Phys. Lett. {\bf 74B} (1978) 61.}
\lref\GHL{J. P. Gauntlett, J. A. Harvey and J. Liu,
Nucl. Phys. {\bf B409} (1993) 363.}
\lref\harr{B. Harrington and H. Shepard, Phys. Rev. {\bf D17} (1978) 2122.}
\lref\rossi{P. Rossi, Nucl. Phys. {\bf B149} (1979) 170.}
\lref\hetsol{A. Strominger, Nucl. Phys. {\bf B343} (1990) 167;
E: Nucl. Phys. {\bf B353} (1991) 565.}
\lref\MO{C. Montonen and D. Olive, Phys. Lett. {\bf 72B} (1977) 117. }
\lref\osborn{H. Osborn, Phys. Lett. {\bf 83B} (1979) 321.}
\lref\gpy{D. J. Gross, R. D. Pisarski and L. G. Yaffe, Rev. Mod. Phys.
{\bf 53} (1981) 43.}
\lref\fromzum{B. Zumino, Phys. Lett. {\bf B69}, (1977) 369.}
\lref\don{
J. A. Harvey and A. Strominger, Comm. Math. Phys. {\bf 151} (1993) 221. }
\lref\smon{J. P. Gauntlett,
Nucl. Phys. {\bf B400} (1993) 103; Nucl. Phys. {\bf B411} (1993) 443.}
\lref\blum{J. Blum, Phys. Lett. {\bf B333} (1994) 92.}
\lref\grossman{B. Grossman, Phys. Lett. {\bf A61} (1977) 86.}
\lref\dualrefa{A. Font, L. E. Ib\'a\~nez, D. L\"ust and F. Quevedo,
Phys. Lett. {\bf B249} (1990) 35;
S. J. Rey, Phys. Rev. {\bf D43}  (1991) 526.}
\lref\dualrefb{
A. Sen, Nucl. Phys. {\bf B404} (1993) 109; Phys. Lett. {\bf B303} (1993) 22;
Mod. Phys. Lett. {\bf A8} (1993) 2023;
J. H. Schwarz and A. Sen, Nucl. Phys. {\bf B411}
(1994) 35.}
\lref\inprep{J. P. Gauntlett and J. A. Harvey, work in progress}
\lref\HM{J. Horne and G. Moore, ``Chaotic Coupling Constants,'' preprint
YCTP-P2-94, RU-94-25 (hep-th/9403058)}
\lref\witten{E. Witten, Nucl. Phys. {\bf B202} (1982) 253.}
\lref\orbi{L. Dixon, J. A. Harvey, C. Vafa and E. Witten, Nucl. Phys. {\bf
B261} (1985) 678;
Nucl. Phys. {\bf B274} (1986) 285. }
\lref\tod{A. Todorov, Inv. Math. {\bf 61} (1980) 25.}
\lref\nati{N. Seiberg, Nucl. Phys. {\bf B303} (1988) 286.}
\lref\rohmw{R. Rohm and E. Witten, Ann. Phys. {\bf 170} (1986) 454.}
 \lref\bddf{T. Banks, M. Dine, H. Dijkstra and W. Fischler,
Phys. Lett. {\bf B212} (1988) 45.}
\lref\sgrav{E. Cremmer, J. Scherk and S. Ferrara, Phys. Lett. {\bf 74B} (1978)
61 \semi
E. Bergshoeff, M. de Roo and B. de Wit, Nucl. Phys. {\bf B182} (1981) 173 \semi
M. K. Gaillard and B. Zumino, Nucl. Phys. {\bf B193} (1981) 221 \semi
M. de Roo, Nucl. Phys. {\bf B255} (1985) 515.}
\lref\kkref{R. Sorkin, Phys. Rev. Lett. {\bf 51} (1983) 87 \semi
D. Gross and M. Perry, Nucl. Phys. {\bf B226} (1983) 29.}
\lref\narain{K. S. Narain, Phys. Lett. {\bf B169} (1986) 41.}
\lref\nsw{K. S. Narain, M. H. Sarmadi, and E. Witten, Nucl. Phys. {\bf B279}
(1987) 369.}
\lref\stw{A. Shapere, S. Trivedi and F. Wilczek, Mod. Phys. Lett. {\bf A6}
(1991) 2677.}
\lref\jm{C. V. Johnson and R. C. Myers, preprint hep-th/9406069. }
\lref\atiyah{M. F. Atiyah, ``Geometry of Yang-Mills Fields'' in Lezioni
Fermiane, Accademia Nazionale dei Lincei Scuola Normale Superiore,
Pisa (1979). }
\lref\adhm{M. F. Atiyah, N. J. Hitchin, V. G. Drinfeld and Yu. I. Manin,
Phys. Lett. {\bf 65A} (1978) 425.}
\lref\ah{M.F. Atiyah and N. J. Hitchin, The Geometry and Dynamics of
Magnetic Monopoles, Princeton University Press, 1988.}
\lref\thooft{G. 't Hooft, Nucl. Phys. {\bf B153} (1979) 141 ; Acta Phys.
Austriaca Suppl.
{\bf XXII} (1980) 531.}
\lref\wittwist{E. Witten, Nucl. Phys. {\bf B202} (1982) 253.}
\lref\hofftwo{ G. 't Hooft, Commun. Math. Phys. {\bf 81} (1981) 267.}
\lref\modrefs{This result is implicit in the work of G. 't Hooft, Phys. Rev.
{\bf D14} (1976) 3432,
see also C. Bernard, Phys. Rev. {\bf D19} (1979) 3013. }
\lref\unpub{J. P. Gauntlett and J. A. Harvey, unpublished.}
\lref\nfourref{S. Mandelstam, Nucl. Phys. {\bf B213} (1983) 149 \semi
L. Brink, O. Lindgren and B. Nilsson, Phys. Lett. {\bf 123B}(1983) 323 \semi
P. S. Howe, K. S. Stelle and P. K. Townsend, Nucl. Phys. {\bf B214} (1983) 519
\semi
P. S. Howe, K. S. Stelle and P. K. Townsend, Phys. Lett. {\bf 124B} (1983) 55
\semi
E. Martinec, Phys. Lett. {\bf B171} (1986) 189.}
\lref\wbrane{C. Callan, J. A. Harvey and A. Strominger, Nucl. Phys. {\bf B359}
(1991) 40.}
\lref\dufflu{M. Duff and J. Lu, Nucl. Phys, {\bf B354} (1991) 129; {\bf B357}
(1991) 129.}
\lref\dufflutwo{M. Duff and J. Lu, Phys. Rev. Lett. {\bf 66} (1991) 1402.}
\lref\bmt{P. J. Braam, A. Maciocia, and A. Todorov, Invent. math. {\bf 108}
(1992) 419.}
\lref\kmir{F. Bogomolov and P. J. Braam,
Commun. Math. Phys. {\bf 143} (1992) 641.}
\lref\hoffmod{G. 't Hooft, Phys. Rev. {\bf D14} (1976) 3432.}
\lref\dist{J. Distler and S. Kachru, PUPT-1464, hep-th@xxx/9406091.}
\lref\bhrefs{recent refs on bh, Giddings, Polchinski, Strominger,
Kallosh, Duff???, what else?}
\lref\coleman{S. Coleman, ``Classical Lumps and their Quantum Descendants'' in
Aspects of Symmetry, Cambridge University Press, 1985.}
\lref\moore{G. Moore, hep-th/9305139.}
\lref\gir{L. Girardello, A. Giveon, M. Porrati, and A. Zaffroni,
hep-th/9406128.}
\lref\swb{N. Seiberg and E. Witten, Nucl. Phys. {\bf B431}
(1994) 484; hep-th/9408099.}
\lref\swa{N. Seiberg and E. Witten, Nucl. Phys. {\bf B426} (1994) 19;
ERRATUM-ibid. {\bf B430} (1994) 485;
hep-th/9407087.
}
\lref\dualrefs{C. Montonen and D. Olive, Phys. Lett. {\bf 72B} (1977)
117;
P. Goddard, J. Nuyts and D. Olive Nucl. Phys. {\bf B125} (1977) 1.}
\lref\nati{N. Seiberg,  Nucl. Phys. {\bf B435} (1995) 129;
hep-th/9411149; K. Intriligator and N. Seiberg,  hep-th/9503179}
\lref\hitch{N. Hitchin as quoted in \ms . }
\lref\wittenth{E. Witten, Phys. Lett. {\bf B86} (1979) 283. }
%
%-------------------
% title page
%-------------------
%
\Title{\vbox{\baselineskip12pt
\hbox{EFI-95-56}
\hbox{CALT-68-2017}
\hbox{hep-th/9508156}}}
{\vbox{\centerline{$S$-Duality and the Dyon Spectrum in }
\centerline{$N=2$ Super Yang-Mills Theory}}}
{
\baselineskip=12pt
\bigskip
\centerline{Jerome P. Gauntlett}
\bigskip
\centerline{\sl Lauritsen Lab}
\centerline{\sl California Institute of Technology}
\centerline{\sl Pasadena, CA 91125}
\centerline{\it jerome@theory.caltech.edu }
\bigskip
\centerline{Jeffrey A. Harvey}
\bigskip
\centerline{\sl Enrico Fermi Institute, University of Chicago}
\centerline{\sl 5640 Ellis Avenue, Chicago, IL 60637 }
\centerline{\it  harvey@poincare.uchicago.edu}

\bigskip

\centerline{\bf Abstract}
We study the dyon spectrum in $N=2$ super Yang-Mills theory with
gauge group $SU(2)$
coupled to $N_f$ matter multiplets in the fundamental representation.
{}For magnetic charge one and two we determine the spectrum explicitly
and show that it is in agreement with the duality predictions of Seiberg and
Witten. We briefly discuss the extension to higher
charge monopoles for the self-dual $N_f=4$ case and argue that the conjectured
spectrum of dyons predicts
the existence of certain harmonic spinors on the moduli space of higher
charge monopoles.

\vskip 1.5in
\noindent

}

%\draftmode

%\Date{April, 1995}
%
%----------------------
% Body of Paper
%----------------------

\newsec{Introduction}

There has been impressive progress recently in understanding
the dynamics of $N=2$
supersymmetric gauge theories \refs{\swa, \swb}.
This progress has relied
on the powerful constraints  which come from the
holomorphic structure of supersymmetric gauge theories and on
a duality between electric and magnetic degrees of freedom \dualrefs.  This
duality symmetry is
far from  apparent
in the current formulation of gauge theory and is still rather poorly
understood.

It has also become clear that there are theories which may possess
an exact duality, relating correlation functions in the dual theories
at all scales, and other theories which possess only effective dualities
in the infrared limit.  It may be that a full understanding of theories
which possess an exact duality will lead to a better understanding
of theories with effective duality.  One class of theories that are
thought to possess an exact duality are the $N=4$ supersymmetric Yang-Mills
theories \refs{\osborn,\senb}.
These theories have perturbatively vanishing  beta functions.
In \swb\ it was conjectured that the simplest $N=2$ gauge theory with
perturbatively vanishing beta function i.e. gauge group $G=SU(2)$ and $N_f=4$
hypermultiplets in the fundamental representation, may also
have an exact $SL(2,Z)$ electric-magnetic $S$-duality
\foot{The precise duality group involves a semi-direct product
of $SL(2,Z)$ and
$Spin(8)$ which is described in \swb\
and section 2 below.}. The arguments of \swb\ were
based on an analysis of the dynamics of
the theory and its relation to the dynamics of theories with $N_f< 4$.

It was further noted in \swb\ that $S$-duality also makes predictions about
the spectrum of  BPS dyon states in the $N_f=4$ theory.
These are states whose masses saturate a Bogomol'nyi bound depending on
their electric and magnetic charges.
In this paper we will analyze the spectrum of BPS dyons
using semi-classical techniques and translate the conjectured spectrum
into predictions about the existence of certain harmonic forms on
the moduli space of BPS monopole solutions. We will
verify these predictions in the one and two-monopole
sectors of the theory.  The analysis is similar in spirit to that
of Sen \senb\ who made and verified the analogous prediction in
the context of $N=4$ super Yang-Mills theory. Our analysis
for $N_f=4$ also verifies, {\it en passant},
a prediction made in \swb\ for
the $N_f=3$ theory concerning the existence of a dyon state
with magnetic charge two.

The outline of this paper is as follows. In section 2 we summarize
the structure of $N=2$ super Yang-Mills theory and the results and predictions
of \swb\ that
are relevant to the problem at hand. In section 3 we discuss monopole
dynamics in $N=2$ gauge theories emphasising the effects
of the coupling to matter fermions. In particular we show that
the low-energy monopole dynamics is determined by a
supersymmetric
quantum mechanics on the moduli space of BPS $k$-monopole configurations
coupled to a natural $O(k)$ connection that is constructed from the
matter fermion zero modes.
Our main results are in section 4
which contains an analysis of the BPS dyon
spectrum in the sectors with magnetic charge one and two.  The analysis
in the charge two sector
eventually reduces to a calculation of the index of the Dirac operator
on the Atiyah-Hitchin manifold coupled to the Levi-Civita connection
and an additional $O(2)$ connection.
We show that  for magnetic charge one and two the dyon spectrum is in
agreement with the conjectures of \swa.
We also discuss the extension of these conjectures to
higher monopole moduli spaces.  The final section contains brief conclusions.

\newsec{$N=2$ Super Yang-Mills and the Seiberg-Witten Conjecture}
\subsec{$N=2$ Without Matter}

Pure $N=2$ super Yang-Mills theory with gauge group $SU(2)$ involves a
single vector supermultiplet consisting of a gauge field $A_\mu$, Weyl
fermions $\lambda$, $\psi$, and a complex scalar $\phi$, all in the
adjoint representation.  The fields $A_\mu , \lambda$ comprise
an $N=1$ vector multiplet $W_\alpha $ while $ \psi , \phi $ are the
components of a $N=1$ chiral superfield $\Phi $.
In component form the bosonic part of the Lagrangian is
\eqn\one{S = {1 \over 16 \pi} {\rm Im}  \int  \tau  {\rm Tr}
(F \wedge F + i \ast F \wedge F ) +
{1 \over g^2} \int d^4x \left( (D_m \phi)^{\dagger} (D^m \phi) -
[\phi,\phi^{\dagger} ]^2  \right)  ,}
where $\tau = \theta/2 \pi + i 4 \pi /g^2$.

\subsec{Classical Theory}

Classically the theory defined by \one\ has a vacuum state for every
gauge inequivalent  minimum of the potential
\eqn\two{ V(\phi) = {1 \over g^2} {\rm Tr} [\phi,\phi^{\dagger} ]^2 .}
If we choose a gauge in which $\phi = {1 \over 2} a \sigma^3$, the classical
moduli space of vacua is the
complex $u$ plane with $u = {1 \over 2} a^2 = {\rm Tr} \phi^2$.

{}For each point with $u \ne 0$ the perturbative spectrum
of the theory consists of
the massless photon and its superpartners and the massive $W^\pm$ states
and their superpartners.
At the semi-classical level, the spectrum also includes monopole
and dyon states with spin $0$ and $1/2$ with charges
$(n_m,n_e)=(1,n_e)$,
where the two integers $n_m$ and $n_e$ specify the electric
and magnetic charges of the state, respectively \refs{\osborn,\gaunt}.
After incorporating the corresponding
antiparticles, these dyons fill out an $N=2$ hypermultiplet of
the $N=2$ supersymmetry.

The masses of all states in this theory, including monopoles
and dyons, satisfy a Bogomol'nyi bound
\eqn\three{ M \ge \sqrt{2} |Z| =\sqrt{2}|a(n_e+\tau n_m)|.}
This bound is derived from the $N=2$
supersymmetry algebra in which $Z$ appears as a central charge \WO.
States that saturate the Bogomol'nyi bound are called BPS states and
form short representations of the $N=2$ supersymmetry algebra.

The perturbative electrically charged
states are all BPS states.
{}For non-zero $n_m$, the Bogomol'nyi bound is saturated if and only if
the classical monopole solutions obey the first order
Bogomol'nyi equations
\eqn\four{ B^i = \pm D^i \phi.}
Here the upper sign corresponds to monopoles ($n_m>0$) and the lower
sign to anti-monopoles ($n_m<0$). The moduli space of
solutions of these Bogomol'nyi equations
is the starting point for determining the
semiclassical existence of BPS monopole
and dyon states, as we shall discuss later.

\subsec{Quantum Theory}

The quantum theory is much more complicated. We will not attempt to
summarize the analysis of \swa\ but will just note the following points.

\item{1.} The quantum moduli space ${\cal M}$  is also the $u$-plane but
with singularities at $u= \pm1, \infty$. Over the $u$-plane there is a flat
$SL(2,Z)$ bundle with specified monodromy.
\item{2.} The full $SU(2)$ gauge symmetry is never
restored on ${\cal M}$. As a result
the theory should possess magnetic monopoles at all values of $u$.
\item{3.} Renormalization of the formula \three\
and the monodromy are consistent
with dyons of charge $(n_m,n_e)= (1,n_e)$ becoming
massless at the singularities
at $u = \pm 1$.
\item{4.} Although there are jumping curves \swa\ across which BPS
states can decay, it is possible to move in from weak coupling ($u= \infty$)
without crossing such a curve. As a result, the states which become massless
at the singularities $u=\pm 1$ must be visible as BPS states at
weak coupling.

\noindent As we mentioned, and will be reviewed briefly below,
the states $(1,n_e)$ do exist as BPS states at weak coupling.

\subsec{$N=2$ With Matter Fields}

Theories with $N=2$ supersymmetry can also contain hypermultiplets
which consist of two Weyl fermions in conjugate representations of the
gauge group and two complex scalars, also in conjugate representations.
These fields are assembled into $N=1$ chiral superfields $Q$ and $\tilde Q$
transforming in conjugate representations.
{}Following \swb\ we will consider $N_f$ hypermultiplets in the fundamental
representation of $SU(2)$, $Q^I$, $ {\tilde Q}_I $, $I = 1,2, \ldots N_f $.

The terms in the Lagrangian involving the hypermultiplets consist
of canonical kinetic energy terms as well as a coupling term
given in $N=1$ superfield language by the superpotential
\eqn\six{W = \sqrt{2} \sum_I {\tilde Q}_I \Phi Q^I. }
The analysis of \swb\ relied on the possibility of adding
mass terms  for the hypermultiplets in \six. Here we will be content
to analyze the theory in the limit of vanishing masses for these fields.

The global flavor symmetry group of
\six\ will be important in the later analysis.
{}For general gauge group there is a $SU(N_f) \times U(1) $ flavor symmetry
which acts on the fields $Q^I$, $ {\tilde Q}_I $ which transform as
a $N_f, {\bar N}_f $. However for $SU(2)$ the fundamental representation
is pseudoreal rather than complex and as a result $Q^I$ and ${\tilde Q}_I$
lie in equivalent representations. This leads to a  $SO(2N_f)$ flavor
symmetry which can be made evident through a change  of basis:
$(Q^I,{\tilde Q}_I)\to Q^{'i}$,$i=1,\dots 2N_f$. As we shall see,
there exist monopoles
and dyons transforming as spinors of $SO(2N_f)$ and so more precisely
the flavor symmetry is $Spin(2N_f)$.

\subsec{Classical Theory}

The classical moduli space of vacua of \six\  is complicated by the
possibility of the matter scalar fields acquiring vacuum expectation values.
This leads to ``Higgs branches'' of the classical moduli space along
which the gauge symmetry is completely broken. Since these branches
do not have classical monopole solutions we will not consider them further.
There is in addition a ``Coulomb branch''  along which the gauge
symmetry is broken to $U(1)$. This branch can again be parametrized by the
gauge invariant quantity $u= \langle {\rm Tr} \phi^2 \rangle$.
Just as in the pure $N=2$ case, on the Coulomb branch the masses
of all states in the theory
satisfy  the Bogomol'nyi
bound \three. The perturbative states saturate the bound and hence are
BPS states. We will investigate in detail the existence
of additional magnetically charged BPS states in later sections.

\subsec{Quantum Theory}
The structure of the quantum theory is now considerably more complicated
but again involves an analysis of the monodromy of families of elliptic curves.
{}For our purposes the main points are the following.

\item{1.} BPS states which become massless at the singularities in the
$u$ plane must again be visible as BPS states at weak coupling.
\item{2.} The singularities for $N_f=1,2$ are consistent with dyons of
charge $(n_m,n_e)=(1,n_e)$ becoming massless.
\item{3.} The singularities for $N_f=3$ require a state $(2,1)$
that transforms as an $SO(6)$ singlet to become
massless. The existence of this BPS state at weak coupling is
thus required for consistency of the analysis in \swb.
\item{4.} The $N_f=4$ theory is a scale invariant theory
with, for vanishing bare masses, no renormalization
of the BPS mass formula \three. The quantum moduli space
for this theory is thus the same as the classical moduli space.
In particular the $SU(2)$ symmetry is restored at the origin of the
$u$ plane.
There is a conjectured exact $S$-duality which predicts the presence
of a $SL(2,Z)$ invariant dyon spectrum as discussed in the following
subsection.

\subsec{Predictions for the Dyon Spectrum}

As mentioned above, analysis of the $N_f=3$ theory predicts the existence
of a BPS state with $(n_m,n_e)=(2,1)$ at weak coupling.
The $N_f=4$ theory gives rise to a richer set of
predictions as a consequence
of a conjectured exact $S$-duality of the spectrum.  The
precise duality group is conjectured to be the semi-direct product
$Spin(8)\semidirect
SL(2,Z)$ \swb. The mod 2 reduction of $SL(2,Z)$ is homomorphic
to $S_3$, which is both the permutation group of three objects and
the group of outer automorphisms of $Spin(8)$. Thus $SL(2,Z)$ acts on
$Spin(8)$ via this homomorphism.

The $SL(2,Z)$ action can be made
more explicit as follows. Label the states by $(n_m, n_e)_r$
where $n_m,n_e$ are the magnetic and electric charges,
respectively and $r$ denotes its $Spin(8)$ representation.
The action of the $SL(2,Z)$
matrix
\eqn\matxy{ M=\pmatrix{\alpha&\beta\cr\gamma&\delta \cr} }
is then given by
\eqn\slt{
\eqalign{\tau&\to{\alpha\tau+\beta\over\gamma\tau+\delta}\cr
a&\to(\gamma\tau+\delta)^{-1}a\cr
(n_m,n_e)_r&\to[(n_m,n_e)M^{-1}]_{r^{'}}.\cr}
}
The representation $r^{'}$ is determined by triality. The vector ($v$),
spinor ($s$) and conjugate spinor ($c$) representations $r$ are
transformed via the $SL(2,Z)\to S_3$ homomorphism. Explicitly,
the mod 2 reduction of the $SL(2,Z)$ matrix gives the following
permutations:
\eqn\hom{
\eqalign{
\left(\matrix{0&1\cr1&0\cr}\right)&\to\Bigg\{\matrix{
v&\to&s\cr s&\to&v\cr c&\to&c \cr}\cr
\left(\matrix{1&0\cr 1&1\cr}\right)&\to\Bigg\{\matrix{
v&\to&c\cr s&\to&s\cr c&\to&v \cr}\qquad\qquad etc.\cr}
}

Beginning with the hypermultiplet of states
$(0,1)$ in the ${\bf 8_v}$ representation,
the elementary quark multiplet,
the $SL(2,Z)$ action generates the orbit of states $(p,q)$ with
$p$ and $q$ relatively prime, with the $Spin(8)$ representation
determined by the mod 2 grading:
\eqn\assig{
(0,1)-{\bf 8_v};\qquad (1,0)-{\bf 8_s}; \qquad (1,1)-{\bf 8_c}.}
Returning to the Bogomol'nyi bound \three\ and using the triangle
inequality we deduce that these states have mass strictly less than
that of any possible  decay products  and hence must be stable. Thus,
duality predicts that this orbit of states exist
in these specific representations, that they saturate the
Bogomol'nyi bound and that they form hypermultiplets of the
underlying $N=2$ supersymmetry.

Let us now consider the vector multiplet $(0,2)$ containing
the $W$ boson which transforms as a singlet under $Spin(8)$.
{}From the Bogomol'nyi bound we first note that
the $W$ boson is only neutrally stable to the decay into
two quarks. Seiberg and Witten suggested that it is possible
that the $W$ bosons are in fact distinct states somewhat
analogous to bound states at threshold in non-relativistic
quantum mechanics. If this is the right interpretation,
and we will in fact provide evidence that it is, self-duality
predicts another orbit of states $(2p,2q)$ with $p$ and $q$
relatively prime, each at threshold. Like the $W$ boson multiplet,
these states should
all transform as singlets under $Spin(8)$ and fill out a vector
multiplet.

Thus, starting with the known
states, duality predicts an orbit of hypermultiplets $(p,q)$
transforming in eight-dimensional representations of $Spin(8)$
and possibly an orbit of vector multiplets $(2p,2q)$ transforming as singlets
\foot{Note that, as in the $N=4$ case,
the states $(0,0)$, the photon multiplet,
lie in a single $SL(2,Z)$ orbit. }. Since the quantum moduli
space is the same as the classical moduli space, there are no
jumping curves where BPS states can decay. Consequently, the
predicted orbits should exist for all points in the moduli space
and in particular for weak coupling where we can look for them using
semi-classical techniques.

{}For monopole number 1, duality requires a tower of hypermultiplets
$(1,2q)$ transforming
as ${\bf 8_s}$ and another $(1,2q+1)$ transforming as ${\bf 8_c}$.
{}For monopole
number 2, a tower of hypermultiplets $(2,2q+1)$ transforming
as ${\bf 8_v}$ is required. In addition, if the interpretation of the
$W$ boson mentioned above is correct, a tower of vector multiplets
$(2,2q)$ transforming as $Spin(8)$ singlets is needed.
We will  provide  evidence for the existence
of all of these states in the ensuing sections.

\newsec{ Monopole Dynamics in the Moduli Space Approximation}
At weak coupling, the dynamics of monopoles
can be studied using
semiclassical techniques. In this section we
show how the low-energy dynamics and in particular
the spectrum can be determined by analyzing a particular
supersymmetric quantum mechanics.

\subsec{Monopole moduli space}
Working in the $A_0=0$ gauge, static
monopole solutions to the classical
equations of motion are
given by solutions to the Bogomol'nyi equations \four.
The monopole moduli space is defined as
the set of gauge equivalence classes of solutions. {}For monopole
charge 1 the moduli space is simply ${\cal M}_1=R^3\times S^1$; the $R^3$
corresponds to the location of the monopole in space and the
phase $S^1$ corresponds to electric charge:
it arises from ``large" gauge transformations on the monopole
solution (this is the way dyons arise in the $A_0=0$ gauge).

The general $k$-monopole moduli space, ${\cal M}_k$,
is a
$4k$-dimensional hyperK\"ahler manifold.
It is
possible to separate the center of mass motion and the total
electric charge and consequently there is an isometric decomposition
\eqn\mod{
{\cal M}_k=R^3\times{S^1\times \tilde{\cal M}^0_k\over Z_k},}
where $\tilde{\cal M}^0_k$ is a $(4k-4)$-dimensional,
simply connected hyperK\"ahler manifold.

\subsec{Fermion zero modes}
Differentiating the most general BPS monopole solution
with respect to the moduli gives the bosonic
zero modes in the small fluctuations about the solution.
With fermions present there are additional fermionic zero modes.
Let us first consider the adjoint fermions. The Callias index
theorem \callias\ implies that for monopole
number $k$ the two Weyl fermions in the adjoint will
give rise to $2k$ complex zero modes.
Since the BPS monopole solutions saturate the Bogomol'nyi bound
\three, the solutions break half of the supersymmetries.
The broken supersymmetries acting on the bosonic solution
lead to four fermion Goldstone modes. {}For a single monopole
these are the only zero modes that come from the adjoint fermions.
{}For higher monopole number, one
can show that the bosonic and fermionic zero modes are paired
by the unbroken supersymmetries \gaunt.
Since the bosonic zero modes are simply tangent vectors of ${\cal M}_k$,
the set of adjoint fermionic zero modes naturally leads to the tangent bundle.
Using the language of \ms\ we can say that the index bundle of
the gauge equivalence classes of Dirac operators in the adjoint representation
parametrized by
points on the moduli space is simply the tangent bundle.

Now let us consider the fermion zero modes arising from
the matter fermions in the  fundamental representation.
The Callias index theorem states that
for monopole number $k$ there are $k$ real zero modes for
each fundamental Weyl fermion. These zero modes are not related to
any bosonic zero modes by supersymmetry.
{}For a single Weyl fermion
Manton and Schroers discuss the index bundle of the
Dirac operators in the fundamental representation and show that it is an
$O(k)$ bundle over the moduli space ${\cal M}_k$ \ms.
They denote it ${\rm Ind}_k$.
This bundle has a natural connection
given by
\eqn\con{
A_a^{AB}(X)= \int d^3x \lambda^A(x,X)^\dagger {\p\over\p X^a}
 \lambda^B(x,X),
}
where $\lambda^A(x,X^a)$, $A=1,\dots,k$ are the zero modes
around a monopole solution
specified by the coordinates $X^a$ on ${\cal M}_k$.
The curvature of this connection is of type $(1,1)$ with respect
to each of the three complex structures on ${\cal M}_k $ and
hence the curvature is anti-self-dual \hitch.
Note that the analogue of the connection \con\
for the index bundle of the adjoint fermions
is the Levi-Civita connection on ${\cal M}_k$.
{}For the $N=2$ models we are interested in, we have $N_f$
hypermultiplets and hence $2N_f$ Weyl fermions in the fundamental
representation. This gives rise
to $2kN_f$ fermion zero modes which leads to $2N_f$ copies
of the $O(k)$ bundle.

\subsec{Collective coordinate expansion}

Let us first briefly summarize how the collective coordinate expansion works
for pure N=2 QCD \gaunt. As usual, for each zero mode one introduces
a collective coordinate. {}For the bosonic zero modes these
are just the coordinates on the moduli space,
the moduli $X^a$ themselves. {}For the $2k$ complex
fermionic zero modes arising from
the adjoint fermions, one
must introduce $4k$ real Grassmann-odd collective coordinates $\psi^a$. We
expect the low-energy dynamics to be dominated by the dynamics of
the zero modes. We can encapsulate this in a low-energy
ansatz for the fields by assuming that all time
dependence is via the collective coordinates. Heuristically, we have
\eqn\ansatz{
\eqalign{A_i(x,t)&=A_i(x,X^a(t)),
\qquad\Phi(x,t)=\Phi(x,X^a(t)),\cr
\psi(x,t)&=F(\delta_aA_i,\delta_a\Phi,\psi^a(t)).\cr}
}
Here $(\delta_aA_i,\delta_a\Phi)$, $a=1,\dots,4k$,
are the bosonic zero modes and
$F$ is the functional determining the supersymmetric pairing
between the bosonic and fermionic zero modes mentioned in
the last subsection (the explicit form of \ansatz\ is given
in \gaunt). Substituting this ansatz into the pure $N=2$ QCD Lagrangian
and integrating over space then leads to an $N=2$ supersymmetric
quantum mechanics on the moduli space of BPS monopoles:
\eqn\sqmp{
S={1\over 2}\int dt {\cal G}_{ab}(X)(\dot X^a \dot X^b + i\psi^a D_t \psi^b),
}
where ${\cal G}_{ab}$ is the metric on the moduli space which arises
from the kinetic energy terms in the field theory \manton.

With the inclusion of hypermultiplets
we must introduce
more Grassmann-odd collective coordinates corresponding
to the extra
fermionic zero modes.
The
low energy ansatz will include the terms
\eqn\ansatzm{
\lambda^i(x,t)={1\over {\sqrt 2}}\sum_{A}\rho^{iA}(t)\lambda^A(x, X^a(t)),
}
where $\lambda^A(x, X^a)$, $A=1,\dots , k$
are the fermion zero modes introduced in \con\
and $\rho^{iA}(t)$, $i=1,\dots,2N_f$ are the
real Grassmann-odd collective coordinates.
Substituting into the Lagrangian and integrating over space, yields for
the kinetic term:
\eqn\kin{
\eqalign{\int d^4x \bar\lambda^i\Dsl\lambda^i
&\to {1\over 2}\int d^4x\left[\rho^{iA}(t) \lambda^A(x,X(t))
^\dagger\right] \p_t
\left[\rho^{iB}(t)
\lambda^B(x,X(t))\right]\cr
& ={1\over 2} \int dt(\rho^{iA} \dot\rho^{iA} + \rho^{iA} A^{AB}_a \dot
X^a \rho^{iB})\cr
&\equiv {1\over 2}\int dt \rho^{iA} {\cal D}_t \rho^{iA},\cr }
}
where in the first line we used the fact that $\lambda^A$ are zero modes
and in the second we have used \con\ and have chosen a basis
of zero modes satisfying
\eqn\cont{\int d^3x \lambda^A(x,X)^\dagger\lambda^B(x,X)=\delta^{AB}.}

Thus by substituting the full low-energy ansatz \ansatz\ and \ansatzm\
into the Lagrangian we
are led to consider the following supersymmetric quantum mechanics:
\eqn\sqm{
S={1\over 2}\int dt \left({\cal G}_{ab}\left[\dot X^a \dot X^b
+ i\psi^a D_t \psi^b\right]
+i\rho^{iA}{\cal D}_t \rho^{iA} + {1\over 2}F_{ab}^{AB}\psi^a\psi^b
\rho^{iA}\rho^{iB}\right).
}

While we have not provided a detailed derivation of \sqm,
supersymmetry
essentially dictates this result given the presence of the $O(k)$ connection
\con\
in the kinetic term for the matter zero modes \kin. Since the monopole
solutions break half of the supersymmetry, the low energy
quantum mechanics must be invariant under the supersymmetries
arising from the unbroken supersymmetries of the field theory. Given
that the number of real components in the unbroken supersymmetries is
$2\times 4$, we expect to have a quantum mechanics with four real
parameters or $N=2$ supersymmetry.
The action \sqm\ automatically
has $N=1/2$ supersymmetry \ag\
(although there is a mismatch between the number of bosons and fermions this
action still admits supersymmetries that are non-linearly realized). Additional
supersymmetries restrict the target. In particular, for the
action to admit $N=2$ supersymmetry the moduli space must
be hyperK\"ahler and the field strength of the gauge connection must be
$(1,1)$ with respect to each of the complex structures. Happily, both
statements are true.

\subsec{ Supersymmetric Quantum Mechanics}
To discuss the low-energy dynamics of the monopoles and particularly the
spectrum of monopoles and dyons we need to quantize the action \sqm.
The quantization of the action without the matter fermions  $\rho$
and its connection to the spectrum of BPS monopoles
in the pure $N=2$ theory was described in detail in \refs{\gaunttwo,\gaunt}.
The anti-commutation relations
of the $\psi^a$ imply that the states are
either holomorphic forms or spinors on the
moduli space. In fact on a hyperK\"ahler manifold these two descriptions
are equivalent.
We will work with spinors.

The new ingredient in \sqm\ is the presence of the $\rho$ fields.
Setting $\hbar=1$ their commutation relations
are given by
\eqn\ac{
\{\rho^{iA}, \rho^{jB}\}=
%{\hbar\over 2}
\delta^{ij}\delta^{AB}
.}
The monopole states must provide a representation of this
Clifford algebra.  These representations can be decomposed under
$SO(2N_f) \times O(k) $ as will be described in the following section.

The four supersymmetry charges are given by
\eqn\sc{\eqalign{
Q&=\psi^a\pi_a,\qquad Q^{(m)}=\psi^a{J^{(m)b}_a}\pi_b\cr
\pi_a&=p_a-{i\over 4}\omega_{a\hat b\hat c}[\psi^{\hat b},\psi^{\hat c}]
-iA_{a}^{AB}\rho^{iA}\rho^{iB},\cr}}
where $ {J^{(m)b}_a}$ are the three complex structures
and the hatted indices are tangent space indices. Substituting
$p_a=-i\hbar\p_a$ and $\psi^{\hat a}=(2)^{-1/2}\gamma^{\hat a}$
we see that $Q$ is just the Dirac operator acting on
spinors in some representation of $O(k)$.
The Hamiltonian is therefore just the square of this Dirac operator plus
$|a\tau k|$. The additional constant term simply arises from a
topological
boundary contribution that exists in the field theory for solutions
of the Bogomol'nyi equations.

The predictions we are aiming to test involve the existence
of states saturating the Bogomol'nyi bound. Equivalently these
are states that are annihilated by half of the supersymmetry charges
of the $N=2$ field theory.
In the moduli space approximation, this means we should look for states
that are annihilated by the supersymmetry charges in \sc. Moreover,
they should have finite norm as is usual for bound states.
Thus extra magnetically charge BPS states in the spectrum
are in correspondence with the $L^2$ kernel of the
Dirac operator on ${\cal M}_k$ coupled to the $O(k)$ connection
on the index bundle.

\newsec{Dyon Spectrum}
\subsec{Magnetic Charge $1$}
As noted, the moduli space for a single monopoles is $R^3\times S^1$.
The supersymmetric quantum mechanics has a free Hamiltonian.
As for ordinary monopoles, quantization of the bosonic coordinates on
$R^3 \times S^1 $ leads to a spectrum of dyons $(1,n_e)$
with continuous spacetime
momentum, the quantized electric charge resulting
from quantization of the $S^1$ part of the moduli space.  Quantization
of the fermionic coordinates tangent to $R^3 \times S^1$ was described
in \osborn\ and \gaunt. The states can be thought of as
four-component spinors
on $R^3 \times S^1$ and correspond to four different states in the
field theory with spin
$0$ and $1/2$. These four states make up a short BPS
multiplet (an irreducible multiplet of the $N=2$ supersymmetry
algebra that saturates the Bogomol'nyi bound has four states \WO).
If we combine these states with similar states that come
from quantizing the anti-monopoles we obtain
a complete hypermultiplet of $N=2$ supersymmetry.

Let us now analyze the
effect of the $\rho$ zero modes. {}For $k=1$ we have the anti-commutation
relations
\eqn\rowanti{ \{\rho^{i}, \rho^{j}\}=
%{1\over 2}
\delta^{ij}.}
As is well known, the representation of this Clifford algebra consists
of the $2^{N_f}$ dimensional spinor representation of $SO(2N_f)$.
This representation is reducible and splits into two irreducible
representations,
both of dimension $2^{2N_f -1}$, under projection by the chirality operator
in $SO(2N_f)$ which following \swb\ we denote by $(-1)^H$.

However this is not quite the end of the story.  Even for $k=1$ there
is still a non-trivial bundle structure. The $O(1)$ connection on the
 index bundle is non-trivial
over the $S^1$ factor and leads to
\eqn\moby{ {\rm Ind}_1 = R^3 \times {\rm M{\mathaccent"7F o}b }, }
where M\"ob is the M\"obius bundle over $S^1$.
Physically this
arises because the gauge transformation which generates a $2 \pi $
rotation  about the $S^1$ factor acts as the non-trivial element of
the center of $SU(2)$ \refs{\wittenth, \swb}.  Since the $N_f$ matter fermions
transform in the fundamental representation of $SU(2)$, there must
be a correlation between the $U(1)$ charge (as measured by rotation
about the $S^1$ factor) and the $SO(2N_f)$ chirality (as measured by
$(-1)^H$).  Specifically, it is expressed by the constraint
\eqn\swcons{e^{i \pi Q} = (-1)^H ,}
where $Q$ is the charge operator. Thus we see that states with
$(-1)^H=1$ have even electric charge while states with $(-1)^H=-1$
carry odd electric charge.

Putting this together eg for the $N_f=4$ case, yields a tower
of hypermultiplets $(1,2n_e)$ in the ${\bf 8_s}$ representation and
another tower $(1,2n_e+1)$ in the ${\bf 8_c}$ representation
in agreement with $Spin(8)\semidirect SL(2,Z)$ duality \swb.

\subsec{Magnetic Charge $2$}
By factoring out the center of mass motion,
the full two monopole moduli space $\m$ can be expressed as
\eqn\cc{\m = R^3 \times \left( {S^1 \times \mtilz \over Z_2 } \right) }
where $\mtilz$ is the four-dimensional Atiyah-Hitchin manifold.
It has a two-fold cover denoted $\mtil $ which can be written as an
isometric product
\eqn\bb{\mtil \cong R^3 \times S^1 \times \mtilz }
where $\mtilz$ is the two-fold cover of the reduced two monopole moduli
space
$\mz$.
$SO(3)$ acts on $\mz$ and has orbits which are either a $RP_2$ (at the bolt)
or $SO(3)/D$ away from the bolt  with $D \cong Z_2 \times Z_2$ the
subgroup of diagonal matrices in $SO(3)$.
We denote the generator of the explicit $Z_2$ that appears
in \cc\ by $I_3$ (see \gm\ for more details).

Let us first briefly recall
the quantization of the supersymmetric quantum
mechanics that arises when there are no $N=2$ matter multiplets i.e.
\sqm\ with no $\rho$ \gaunt. As noted, the quantization of the $\psi^a$
implies
that the states are spinors on $M_2$.
The structure of \cc\ implies that these spinors
are tensor products of spinors
on $R^3\times S^1$ with spinors on $\mtilz$.
The discussion of the spinors on $R^3\times S^1$ is essentially
as for the single monopole case: the dyon states
$(2,n_e)$ are in a short BPS multiplet with spin $0$ and $1/2$.
Combining them with the anti-monopole states then fills
out the spin content of a hypermultiplet.
In the normalisations we are using, for the pure $N=2$ case the
electric charge is always even.
The effect of the $Z_2$ that appears in \cc\
is that dyon states whose electric charge is (is not) a multiple
of four are associated with
spinors on $\mtilz$ that are even (odd) under the action of the $I_3$.

The supersymmetry charge $Q$ acts as the Dirac operator on the spinors.
Consequently, the Hamiltonian is the sum of the
free Hamiltonian on $R^3\times S^1$
and the square of the Dirac operator on $\mtilz$ and the constant topological
term. Thus, to find new BPS states in the spectrum we must
look for zero energy states on $\mtilz$ or equivalently zero modes
of the Dirac operator. We will in fact see that there are no zero modes
of the Dirac operator and consequently no extra BPS states for the pure
$N=2$ case.

Now we consider the matter fermions.
The index bundle ${\rm Ind}_2$ is a real two-dimensional vector bundle over
$M_2$ with structure group $O(2)$
which is described in detail in \ms.  There is an obstruction to
obtaining an orientable bundle on the non-simply connected
manifold $(S^1\times\mtilz)/Z_2$. One obtains
an orientable bundle
${\widetilde {\rm Ind}_2}$ by pulling ${\rm Ind}_2$ back to
$S^1 \times \mtilz$.  {}For now we work with the $U(1)$ bundle ${\widetilde
{\rm Ind}_2}$. Replacing the real Grassmann parameters $\rho^{iA}$
with complex parameters $\rho^{i}$, the anticommutation relations become
those of annihilation and creation operators:
\eqn\aandc{\{\rho^i,\rho^{j*}\}=\delta^{ij}.}

The states of the supersymmetric quantum mechanics
are now spinors on $M_2$, $|\Psi\rangle$, on
which the algebra \aandc\ is realized. Starting with a state $|\Psi\rangle$
satisfying $\rho^i |\Psi\rangle=0$ we can build up the $\rho$ {}Fock space
by acting
with the $\rho^{i*}$ in the usual manner. There is a correlation between
the number of $\rho^{i*}$'s excited , $N_\rho$, and the $U(1)$ charge carried
by the corresponding spinor on $M_2$. To see this
note that
the supersymmetry charge \sc\ acting on these states takes the form
\eqn\sct{Q=\Dsl-i(N_\rho -N_0)A,}
where $A$ is the $U(1)$ gauge connection and $N_0$ is a normal
ordering constant. This normal ordering constant can be fixed
by a discrete charge conjugation symmetry: for each state with $U(1)$ charge
$q$ there should be another state with $U(1)$ charge $-q$ \jr.
More precisely,
there is a discrete symmetry which combines parity and a ``magnetic charge
conjugation '' described in \ms\ which changes the sign of the $U(1)$
charge and also the electric charge but leaves the magnetic charge
invariant. This fixes
$N_0$ to be $N_0=-N_f$.

Since the $\rho^i$
carry the ${\bf 2N_f}$ representation of $SO(2N_f)$, there is also a
correlation between the $SO(2N_f)$ representation carried by the
state and $N_\rho$ and consequently the $U(1)$ charge $q$.
{}For the cases of most interest, $N_f=4,3$, we display the explicit
correlation between $q$ and the $SO(8)$, $SO(6)$ representations
in the following table.

\centerline{Table 1}
$$
\vbox{ \settabs 3 \columns
\+   $q$   & $SO(8)$ rep& $SO(6)$ rep. \cr
\+   $\pm0$  & {\bf 35} + {\bf 35}& {\bf 10} + ${\bf \overline{10}}$\cr
\+  $\pm1$   & {\bf 56}& {\bf 15}\cr
\+  $\pm2$  & {\bf 28}& {\bf 6}\cr
\+ $\pm3$   & ${\bf 8_v}$& {\bf 1}\cr
\+ $\pm4$  & {\bf 1}\cr}
$$

As in the pure $N=2$ case the spinors are tensor products of
spinors on $R^3\times S^1$ and spinors on $\mtilz$. The states
on $R^3\times S^1$ lead to spin $0$ and spin $1/2$.
Extra BPS
states arise from states that are annihilated by the Hamiltonian
and consequently the supersymmetry charge. Thus we need to look for zero
modes of the Dirac operator on $\mtilz$ coupled to the $U(1)$ connection
which have finite $L^2$ norm.

We begin by discussing some vanishing theorems.
The square of the Dirac operator
acting on a charge $q$ field
can be expressed as
\eqn\van{\Dsl\Dsl\psi=D^2\psi+{1\over 8} R_{abcd}\gamma^{ab}
\gamma^{cd}\psi+{q^2\over 2}F_{ab}\gamma^{ab}
.}
{}For the case of interest, the manifold is four-dimensional and
both curvatures are anti-self-dual.
Using this we have
\eqn\vant{\Dsl\Dsl\psi=D^2\psi
+{q^2\over 2}F_{ab}\gamma^{ab}{(1+\gamma_5)\over 2}\psi
,}
where $\gamma_5=\gamma_{\hat 1}\gamma_{\hat 2}\gamma_{\hat
3}\gamma_{\hat 4}$.

{}First consider $q=0$.
Multiplying \vant\ by $\psi^\dagger$ and integrating by parts
we deduce that $||\Dsl\psi||^2=
||D_a\psi||^2$. Since the only covariantly constant spinor with finite
$L^2$ norm is zero, we deduce that there are no nontrivial
zero modes of the Dirac operator with finite norm.
This means that there are no extra BPS states in the spectrum of the
$N=2$ theory without matter. In fact this is also a necessary requirement
for the $N=4$ theory to be self-dual. The reason is that
a harmonic
spinor on $\mtilz$ is equivalent to
a harmonic holomorphic differential form on $\mtilz$ and
consequently a BPS state in the $N=4$ theory.
However, Sen showed that $S$-duality predicts that only
a single anti-self-dual harmonic form (which cannot be holomorphic)
should exist on $\mtilz$.

{}For general $q$, by the same reasoning, \vant\ implies that the
only non-trivial Dirac zero modes with finite norm satisfy
$\gamma_5\psi=\psi$.
To calculate the number of these we
use the Atiyah-Patodi-Singer
index theorem.
The index theorem for manifolds with boundary
reads in this case
\eqn\aa{Index(\Dsl) = {1 \over 192 \pi^2}
\int_M \tr R \wedge R - {q^2 \over 8 \pi
^2} \int_M
{}F \wedge F + \int_{\partial M} Q + \hf [ \eta(0) + h_D] ,}
where $q$ is the charge of the fermion under the $U(1) \subset O(2)$ part of
the connection.
Here $Q$ involves Chern-Simons-like contributions as described e.g. in
\egh. $\eta(0)$ is the Dirac $\eta$-invariant of Atiyah-Patodi-Singer
and $h_D$ is the number
of harmonic spinors on the boundary.
The index must be evaluated on a fixed finite boundary
$r_0$ and then one takes the limit $r_0\to \infty$. The index then counts
the difference in the number of harmonic spinors satisfying the
Atiyah-Patodi-Singer boundary conditions (see for example \egh).
As far as we are aware
there is no known
proof that these correspond to $L^2$ boundary conditions, but in all cases
that we know of, they appear to. In particular they do for Taub-NUT space
and since $\mtilz$ asymptotically approaches
Taub-NUT space exponentially in the radial distance (see below)
we assume that they do for the Atiyah-Hitchin manifold.

Before evaluating the various terms in \aa\ we first give some
more details about $\mtilz$.
The explicit metric on $\mtilz$ is known and is given by
\eqn\twnine{ds^2 = f(r)^2 dr^2 + a(r)^2 (\sigma_1^R)^2 + b(r)^2
(\sigma_2^R)^2 + c(r)^2 (\sigma_3^R)^2.}
Here the $\sigma_i^R$ are left-invariant one-forms on $SO(3)=S^3/Z_2$ and the
explicit forms are given,
in the
conventions of appendix A, by
\eqn\thirty{ \eqalign{ \sigma_1^R
& = -\sin \psi d \theta + \cos \psi \sin \theta d \phi \cr
 \sigma_2^R & = \cos \psi d \theta + \sin \psi \sin \theta d \phi \cr
 \sigma_3^R & = d \psi + \cos \theta d \phi \cr }}
with $0 \le \theta \le \pi$, $0 \le \phi \le 2 \pi$, $0 \le \psi < 2 \pi$.
The angles are further restricted under the identification
of the discrete right isometry \gm\ (see appendix A)
\eqn\bird{
(\phi,\theta,\psi)I_x= (\pi+\phi,\pi-\theta,-\psi)
.}
Note that we can equivalently let the range of $\psi$ be $0 \le \psi < 4
\pi$ and then divide out by $I_x$.

We will follow
\gm\ in choosing $f(r)= -b(r)/r$.
The radial functions $a(r)$, $b(r)$ and $c(r)$ are given
explicitly in \ah. Here we only need the asymptotic forms.
Near
$r=\pi$ they take the form
\eqn\asymp{\eqalign{
a(r)&=2(r-\pi)\left\{1-{1\over 4\pi}(r-\pi)\right\}+\dots\cr
b(r)&=\pi\left\{1+{1\over 2\pi}(r-\pi)\right\}+\dots\cr
c(r)&=-\pi\left\{1-{1\over
2\pi}(r-\pi)\right\}+\dots.\cr}}
Introducing appropriate Euler angles, it can be shown that after the
identification by $I_x$ the metric
is smooth near $r=\pi$ and that $r=\pi$ is an $S^2$ or bolt \gm.
Near infinity, $r\to\infty$, the functions take the form
\eqn\asympt{\eqalign{
a(r)&=r\left(1-{2\over r}\right)^{1/2}+\dots\cr
b(r)&=r\left(1-{2\over r}\right)^{1/2}+\dots\cr
c(r)&=-2\left(1-{2\over r}\right)^{-1/2}+\dots,\cr}}
where the neglected terms fall off exponentially with $r$.
Thus the metric approaches Taub-NUT space with negative
mass parameter. Since $a=b$ asymptotically, in addition
to the $SO(3)$ isometries arising from the left action there
is an extra $U(1)$ isometry coming from right actions.
Physically, it corresponds to the fact that the relative
electric charge of two widely separated monopoles
becomes a good quantum number.
Due to the identifications on the Euler angles arising from $I_x$, the topology
of the boundary is $S^3/Z_4$.

The curvature two-form  of the $O(2)$ connection on ${\widetilde {\rm Ind}_2}$
is anti-self-dual and $SO(3)$ invariant and is given explicitly by \ms\
\eqn\senform{F = \alpha(r) \left( d \sigma_1^R - {f a \over bc} dr \wedge
\sigma_1^R \right) .}
where $\alpha(r)$ obeys the ordinary differential equation
\eqn\difff{ {d \alpha \over dr} = - {fa \over bc} \alpha . }
{}From physical arguments it is known that the first Chern number of
${\widetilde {\rm Ind}_2}$
is $\pm 1$. Following \ms\ we fix $c_1 = -1$ for concreteness. This fixes
the normalization of $\alpha(r)$ so that $\alpha(\pi)= 1/2$.
As an aside, note that this harmonic anti-self-dual two form is in fact
the same two-form corresponding to a BPS bound state in the
$N=4$ theory found by Sen \senb.

Let us now evaluate \aa\ for $ \mtilz $.
We first consider the Chern-Simons terms.
It has been explicitly
shown that the Chern-Simons pieces vanish for Taub-NUT space
\egh.
Since the Atiyah-Hitchin metric asymptotically approaches
Taub-NUT space and differs only by terms exponentially
small in $r$, we conclude that the Chern-Simons terms make no contribution
to the index.
Next consider the volume terms.
A
straightforward calculation which we present in appendix B gives
\eqn\elcalc{{1 \over 192 \pi^2}
\int_M \tr R \wedge R - {q^2 \over 8 \pi
^2} \int_M
{}F \wedge F  = {1 \over 6} - {q^2 \over 8} .}

Proceeding to the boundary contributions in \aa,
we need to evaluate $\eta(0)$ and $h_D$ on the $S^3/Z_4$ boundary
of $ \mtilz$.
Now the $\eta$ invariant ,
$\eta(0)$, is defined to be the analytic continuation
to $s=0$ of
\eqn\qq{ \eta(s) = \sum_{\lambda \ne 0}  \lambda^{-s} {\rm sign} \lambda ,}
where the $\lambda$ are the eigenvalues of the Dirac operator on the
boundary coupled to the flat $O(2)$ connection on  ${\widetilde {\rm Ind}_2}$.
We present the details of the calculation in the appendix where we also show
that there are no harmonic spinors on the boundary and hence $h_D=0$.
We obtain
\eqn\here{
\eta_{q} (0)=
 {2 \over 3} + {1 \over 4}( [q+2]_4^2 - 4[q+2]_4),}
where $[q+2]_4=(q+2)$ mod 4.
Combining this result with \elcalc\ we conclude that there
are no zero modes with charge $q=0,\pm1, \pm2$, one zero mode
for $q=\pm3$ and two each for $q=\pm4$.

Let us now check how this fits in with the duality conjectures.
{}For $N_f=4$ we want a tower of states $(2,2n_e+1)$ forming
a hypermultiplet transforming as a ${\bf 8_v}$ of $SO(8)$. In addition
we need another tower of states $(2,2n_e)$ transforming
as $SO(8)$ singlets and filling out a vector multiplet.
{}For $N_f=3$ we want a single hypermultiplet of states
$(2,1)$ transforming as a singlet of $SO(6)$.

First consider the hypermultiplets. Recall that the spin
content of a hypermultiplet is $S_z=(0,0,0,0,\pm 1/2, \pm1/2)$
where $S_z$ is the component of spin along the $z$ axis.
We have noted that the spinor on $R^3\times S^1$ corresponds to
four states in a short BPS multiplet with $S_z=(0,0,\pm 1/2)$.
If these spinors are combined with one zero mode of
the Dirac operator on $\mtilz$ with
zero angular momentum we will obtain a
BPS hypermultiplet after we also include the corresponding
states that come from quantizing the zero modes around
the charge 2 anti-monopole\foot{Note that the moduli space for
the anti-monopole is the same as that of the monopole.}.
If in addition the zero
mode of the Dirac operator has $U(1)$ charge
$q=\pm3$ then according to table 1 the
hypermultiplet will transform as an ${\bf 8_v}$ for the $N_f=4$ case
and as a singlet for the $N_f=3$ case.

Now consider the vector multiplets.
The spin content of the vector multiplet is
$S_z=(0,0,\pm1/2,\pm1/2,\pm1)$. To obtain this
spin content we need to combine the spinor on
$R^3\times S^1$ which has spin $S_z=(0,0,\pm1/2)$
with {\it two} zero modes of the Dirac operator on
$\mtilz$, one with $S_z=1/2$ the other with $S_z=-1/2$.
To form a singlet representation of $SO(8)$ the spinors on
$\mtilz$ must be zero modes of the Dirac operator with charge
$q=\pm4$.

Our analysis in terms of the index of the Dirac operator allows
us to count states but for a detailed check of the duality predictions
we also need to identify the angular momentum and electric charges
of these states. We have not done this; to do so would require either
a more sophisticated use of index theory (especially relating
the indices on the bundles ${\rm Ind}_2$ and $\widetilde {\rm Ind_2}$)
or what would be more
desirable, an explicit construction of the zero modes of the Dirac
operator.  In spite of this it is possible on general grounds to
almost completely determine what the spectrum must be.
Consider first the charge $q = \pm 3 $ zero modes.  The constraint
\swcons\ implies that these states must carry odd electric charge.
In addition, charge conjugation symmetry implies that they carry
opposite electric charge.  If they both carried arbitrary odd electric
charge then we would get twice as many states as required
by duality. However there is in fact a $Z_4$ condition on the
electric charges which follows from the analysis of  \ms. The
$Z_2$ symmetry $I_3$ acting on $M_2$ squares to give a translation
about the $S^1$ factor. But since the holonomy about the $S^1$
factor is $-1$ (for odd $U(1)$ charge ) there is in fact a mod $4$ condition
on the charges.  If the state at $U(1)$ charge $q=-3$ has electric
charge $+1$ $(-1)$ mod 4 then the state at charge $q= +3 $ must have
electric charge $ -1 $ $(+1)$ mod 4 by charge conjugation.   Given this
electric charge assignment and $N=2$ supersymmetry the only
consistent possibility is that this state has angular momentum zero
so that the total sets of states (after including the antiparticles)
transform as hypermultiplets of $N=2$
supersymmetry. For $N_f=4$, we see from table 1 that these states
transform as ${\bf 8_v}$ of $SO(8)$ in complete agreement with
the predictions of \swb. For $N_f=3$ the states transform as singlets
of $SO(6)$ in agreement with the $(2,1)$ state predicted by
\swb.

Now consider the two states we found at $q= \pm 4$.  By \swcons\
these states must carry even electric charge.  An assignment
of charges and spins consistent with duality and charge conjugation
symmetry (which, as mentioned in section 4.2,
actually involves parity as well )
and the presence of a mod 4 condition on the electric charges is the
following.
Assign the two $q=4$ states $S_z = 1/2$, and for one state electric charge
$0$ mod $4$ and the other charge $2$ mod $4$. For the two $q=-4$ states
assign $S_z= -1/2$ and again electric charges $0$ and $2$ mod 4. This
gives a spectrum of charged vector multiplets with a multiplicity consistent
with duality. However, as far as we can see this assignment is not completely
forced by consistency conditions and thus in principle a more refined
test of duality would be provided by an explicit construction of the
zero modes which would allow an unambiguous determination of
these quantum numbers.

\newsec{Magnetic Charge $k>2$}
In order to fully establish the $SL(2,Z)$ duality of the spectrum of
BPS
states in the $N=2$, $N_f=4$ theory it is necessary to generalize the
arguments in the previous section to higher magnetic charge. At
present
this seems difficult without a fuller understanding of the $k$
monopole
moduli space and particularly its asymptotic structure. There has
been
some recent progress in determining this structure for
well-separated
$k$-monopoles \ref\gmnew{G. Gibbons and N. Manton, ``The Moduli space
for
well separated BPS monopoles,'' hep-th/9506052.} .  In the hope that
this may eventually be better understood we will be content here to
state the content of the duality conjecture for $k>2$.

{}For magnetic charge $k$ and $N_f$ flavors the matter fermions have
$2kN_f$ zero modes in the monopole background and the low-energy
effective action involves the coupling of the fermion zero modes
$\rho ^{i A}$, $A=1,2, \ldots k$, $i = 1, 2, \ldots 2N_f $ to an
$O(k)$ connection on the $k$ monopole moduli space ${\cal M}_k $.
The anti-commutation relations of the  $\rho^{iA}$  give rise to a
representation
of the Clifford algebra associated to $O(2kN_f)$ with the zero modes
transforming as the fundamental $2kN_f$-dimensional representation
which decomposes under

\eqn\odecomp{O(2kN_f) \to O(k) \times SO(2 N_f) }
as
$ 2kN_f \to (k,2N_f) $.  The monopole ground state transforms as
the $2^{kN_f}$ dimensional spinor representation of $O(2kN_f)$.

Under the decomposition \odecomp\ we will have
\eqn\spdecom{ 2^{k N_f} \to \sum_i (r^i_k , r^i_{2N_f} ) .}
The actual determination of the irreducible representations of $O(k)$
 that appear in \spdecom\ is somewhat complicated to state in
general.
The important point is that \spdecom\  gives a pairing between
representations
of $O(k)$ and $SO(2N_f)$.  Some general features of this pairing are
immediate and in agreement with the general requirements of duality.
{}For even $k$ we can use the decomposition $O(k) \to SU(k/2) \times
U(1) $
 to write
the Clifford algebra in terms of creation and annihilation operators
which transform as the $2N_f $ of $O(2N_f)$. As a result the ground
state will transform as a sum of tensor representations of $O(2N_f)$
in agreement with the requirements of duality for $N_f=4$. {}For odd $k$,
on the other hand,
we can use the decomposition
$SO(2N_f) \to SU(N_f) \times U(1) $ to write the fermion zero modes
in terms of creation and annihilation operators transforming as
$(N_f,k)+({ \bar N}_f , k ) $ which leads to ground states
transforming
as spinorial representations of $O(2N_f)$ \ref\gsw{See for example
the
discussion in Appendix 5.A of M. Green, J. Schwarz and E. Witten,
``Superstring Theory I, ''  (Cambridge University Press 1987 ).}.

{}For $N_f=4$,  $S$-duality requires an analysis of the index of the
Dirac operator on the monopole moduli space ${\cal M}_k$ coupled
to the $O(k)$ connection on the index bundle $I_k$ through the
representation of $O(k)$ determined by the pairing in \spdecom.
In particular, as in the $k=2$ case analyzed in the last section,
the index for even $k$
should be one  in the $O(k)$ representation paired with the
${\bf 8_v}$ representation of $O(8)$ (representing the $SL(2,Z)$ duals
of the quark hypermultiplets), should be two in the $O(k)$
representation paired with the identity representation of $O(8) $
(representing the $SL(2,Z)$ duals of the gauge boson states) and
should
vanish for all other representations of $O(8)$.
{}For odd $k$, on the other hand, duality predicts index one for the
representation of $O(k)$ paired with the ${\bf 8_s}$ of $O(8)$,
index one for the
representation of $O(k)$ paired with the ${\bf 8_c}$ of $O(8)$ (corresponding
to the $SL(2,Z)$ duals of the hypermultiplets $(1,0)$ and $(1,1)$,
respectively) and
vanishing index for all other representations.
In addition the electric
charge assignments and rotational quantum numbers for these states
must be consistent
with duality.

\newsec{Conclusions}
We have verified the predictions of Seiberg and Witten for the
spectrum of
dyon bound states in $N=2$, $SU(2)$  super Yang-Mills theory coupled to
$N_f$ matter multiplets  in the case of  magnetic charge $k=2$.
The most dramatic result is the existence of a spectrum compatible
with
$SL(2,Z)$ duality for the $N_f =4$ theory as predicted by Seiberg
and Witten.  This includes the existence of the $SL(2,Z)$ duals of
the
gauge bosons as ``bound states at threshold,'' an interpretation
suggested
in \swb\ but not clearly required by duality.

It would be useful to extend the techniques used in this paper
to further check the duality conjecture by giving a more
precise determination of the electric charges of the dyon states
and an explicit calculation
of the angular momentum of the zero modes by
explicit construction of the zero modes of the Dirac operator for
monopole charge 2.
Of most interest of course, would be to extend these
results
to higher magnetic charge by explicitly verifying the predictions
made in section 5.

More generally, it
would be very interesting to extend the duality analysis
of the $SU(2)$, $N_f=4$ case to other gauge groups which
have a field content leading to finite $N=2$ theories.

The analysis we have presented here also makes clear how little is
understood
about duality.
A prediction which is simple to state and follows naturally from an
analysis
of the dynamics of $N=2$ gauge theory can at present be verified only
by detailed calculations involving intricate cancellations and then
only
for low values of the magnetic charge.  Hopefully a deeper
understanding
of duality will allow us to understand the spectrum of dyon bound
states
without recourse to the sort of analysis presented here.

\bigskip
\centerline{\bf Acknowledgements}\nobreak
We thank  A. Dabholkar, F. Dowker,
G. Gibbons, N. Hitchin, G. Moore, J. Preskill,
N. Seiberg, A. Sen,
A. Strominger,  E. Witten and especially S. Dowker for  helpful discussions.
JPG is supported by the U.S.
Dept. of Energy
under Grant No. DE-FG03-92-ER40701
and JH
by NSF Grant No.~PHY 91-23780.

\appendix{A}{Conventions}

The conventions adopted here are essentially the same as in
the \gm\ with some minor changes\foot{They are the same as
{}Fano and Racah \fr\   if we interchange
$\phi$ and $\psi$ and if we replace their $(m',m)$ with $(m,s)$.
They are related to the conventions of  Landau and Lifschitz  \ll\
with the replacements
 $(\alpha,\beta,\gamma) \to (\psi,\theta,\phi)$ and
 $(m',m) \to  (m,s)$.}.
We parametrize the three sphere with Euler angles. The general
$SU(2)$ rotation matrix can be constructed as follows\foot{Note that
this element of $SU(2)$ is denoted $U(\psi,\theta,\phi)$ in \ll.}
\eqn\sthree{
\eqalign{U(\phi,\theta,\psi)=&U_z(\phi) U_y(\theta) U_z(\psi)\cr
&=\left(\matrix{\cos{\theta\over 2}e^{i {(\psi+\phi)\over 2}}&
         \sin{\theta\over 2}e^{-i {(\psi-\phi)\over 2}}\cr
        -\sin{\theta\over 2}e^{i {(\psi-\phi)\over 2}}&
         \cos{\theta\over 2}e^{-i {(\psi+\phi)\over 2}}\cr}\right)
}.}
The ranges of the angles are $0\le\theta\le\pi$, $0\le\phi\le2\pi$
and $0\le\psi< 4\pi$. The $SO(3)$ group manifold is obtained by
restricting the range of $\psi$ to be $0\le\psi< 2\pi$ and identifying
$\psi\sim\psi+2\pi$.

By expanding $U^{-1} dU$ in the basis $({i\over 2}\tau^i)$ where
$(\tau^i)$ are the Pauli matrices, we can construct the left
invariant or ``right" one forms $\sigma^R_i$.
Similarly, the right invariant or ``left" one
forms $\sigma^L_i$ can be constructed from $dU U^{-1}$. We get
\eqn\liof{
\eqalign{\sigma^R_1&= -\sin\psi d\theta+\cos\psi \sin\theta d\phi\cr
\sigma^R_2&= \cos\psi d\theta+\sin\psi \sin\theta d\phi\cr
\sigma^R_3&= d\psi+\cos\theta d\phi\cr}}
and
\eqn\riof{
\eqalign{\sigma^L_1&= \sin\phi d\theta-\cos\phi \sin\theta d\psi\cr
\sigma^L_2&= \cos\phi d\theta+\sin\phi \sin\theta d\psi\cr
\sigma^L_3&= d\phi+\cos\theta d\psi.\cr}}
The superscript $R$ ($L$) refers to the fact that the left (right) invariant
one forms are dual to left (right)
invariant vector fields $\xi^R_i$ ($\xi^L_i$)
which generate right (left) group actions. We will also refer to $\xi^R_i$
as a right vector field and to $\xi^L_i$ as a left vector field.
The explicit form of the dual vector
fields satisfying $\langle\xi^R_i,\sigma^R_j\rangle=\delta_{ij}$ and
$\langle\xi^L_i,\sigma^L_j\rangle=\delta_{ij}$ are given by
\eqn\livf{
\eqalign{
\xi^R_1&= -\cot\theta \cos\psi \p_\psi-\sin\psi \p_\theta+
{\cos\psi\over \sin\theta}
\p_\phi\cr
\xi^R_2&= -\cot\theta \sin\psi \p_\psi+\cos\psi \p_\theta+{\sin\psi\over
\sin\theta}
\p_\phi\cr
\xi^R_3&= \p_\psi\cr}}
and
\eqn\riof{
\eqalign{
\xi^L_1&= -{\cos\phi\over \sin\theta}
\p_\psi +\sin\phi \p_\theta+\cot\theta \cos\phi\p_\phi\cr
\xi^L_2&= {\sin\phi\over \sin\theta}
\p_\psi+\cos\phi \p_\theta-\cot\theta \sin\phi \p_\phi\cr
\xi^L_3&= \p_\phi.\cr}}
%
%(Again note the difference with \gm.)

The left and right invariant one forms satisfy the Maurer-Cartan equations
\eqn\mc{
\eqalign{
d\sigma^R_i&={1\over 2}\epsilon_{ijk}\sigma^R_j\wedge\sigma^R_k\cr
d\sigma^L_i&=-{1\over 2}\epsilon_{ijk}\sigma^L_j\wedge\sigma^L_k.\cr
}}
The Lie brackets of the left and right vector fields are given by
\eqn\lb{
\eqalign{[\xi^R_i,\xi^R_j]&=-\epsilon_{ijk}\xi^R_k\cr
[\xi^L_i,\xi^L_j]&=\epsilon_{ijk}\xi^L_k\cr
[\xi^R_i,\xi^L_j]&=0.\cr}}
The last equation expresses the fact that the right (left) vector fields
are left (right) invariant. Note that the right vector fields
satisfy the same algebra as that of our Lie algebra basis $({i\over 2}\tau^i)$
whereas there is an extra minus sign in the algebra of the left vector
fields.
We define angular momentum operators
\eqn\am{
L^R_i=-i\xi^R_i,\qquad L^L_i=i\xi^L_i
}
satisfying
\eqn\alg{
[L_i,L_j]=i\epsilon_{ijk}L_k
}
for either superscript. We also use the combinations
$L_\pm=L_1\pm i L_2$.

{}Following \fr\ and \ll\ we introduce the Wigner functions\foot{
This is denoted $D^j_{ms}(\psi,\theta,\phi)$ in \ll.}
\eqn\wig{
D^j_{ms}(\phi,\theta,\psi)=e^{im\phi}d^j_{ms}(\theta)e^{is\psi}
}
satisfying
\eqn\ralg{
\eqalign{
L^R_\pm D^j_{ms}&=\left[j(j+1)-s(s\pm1)\right]^{1/2} D^j_{ms\pm1}\cr
L^R_3 D^j_{ms}&=s D^j_{ms}\cr
}}
and
\eqn\lalg{
\eqalign{
L^L_\pm D^j_{ms}&=-\left[j(j+1)-m(m\mp1)\right]^{1/2} D^j_{m\mp 1s}\cr
L^L_3 D^j_{ms}&=-mD^j_{ms}\cr
}}
Note that
$D^{1/2}_{ms}(\phi,\theta,\psi)=U(\phi,\theta,\psi)$.

Next we analyze the discrete symmetries that appear
in the discussion of the two monopole moduli space.
The metric \twnine\ is constructed from $\sigma^R$ and hence is left invariant;
the left vectors
$\xi^L_i$ are Killing vectors. By restricting the range of $\psi$
to be $0\le\psi< 2\pi$ these
generate the isometry group $SO(3)$.
Additional isometries come from right actions.
{}Following \gm\ we consider right actions corresponding to
rotations of $\pi$ about the $x,y$ and $z$ axes.
These $SU(2)$ matrices, consistent with \sthree, are
\eqn\exrot{
U_x(\pi)=\left(\matrix{0&i\cr i&0\cr}\right),\quad
U_y(\pi)=\left(\matrix{0&1\cr -1&0\cr}\right),\quad
U_z(\pi)=\left(\matrix{i&0\cr 0&-i\cr}\right)}
Calculating $U(\phi,\theta,\psi)U_x(\pi)=U(\phi',\theta',\psi')$ etc.
we are led to the following transformations $I_x$ etc. on the angles:
\eqn\here{
\eqalign{
(\phi,\theta,\psi)I_x=& (\pi+\phi,\pi-\theta,-\psi)\cr
(\phi,\theta,\psi)I_y=& (\pi+\phi,\pi-\theta,\pi-\psi)\cr
(\phi,\theta,\psi)I_z=& (\phi,\theta,\pi+\psi).\cr
}}
These transformations each change the sign of
two of the left invariant one forms and hence leave the metric invariant
confirming that they are indeed discrete isometries of the left
invariant metric.
We have chosen the notation in \here\ to emphasize that these
are right actions. Using this notation we have \gm\
\eqn\algeb{
I_x I_z =I_y .}
Note also that $I_x^2=I_y^2=I_z^2$ and is the antipodal map on $S^3$.
Thus, in the definition of the Atiyah-Hitchin manifold one can begin with
left invariant forms on the $SO(3)$ group manifold and then
divide out by $I_x$ or equivalently one can start with $SU(2)$ and
divide
out by $I_x$ (which generates a $Z_4$ in $SU(2)$).
Note the transformation $I_3$ (the $Z_2$ that appears in \cc) is
simply $I_z$ plus an additional action on the coordinate of the $S^1$
(see \gm).

Using the formulae in appendix D of \fr\ or equivalently in section 58
of \ll\ we have  $d^j_{ms}(\pi-\theta)=(-1)^{j-m} d^j_{m-s}(\theta)$ and hence
\eqn\daods{
\eqalign{
D^j_{ms}(gI_x)&=(-1)^jD^j_{m-s}(g)\cr
D^j_{ms}(gI_y)&=(-1)^{j+s}D^j_{m-s}(g)\cr
D^j_{ms}(gI_z)&=(-1)^sD^j_{ms}(g),\cr}}
where $g=(\phi,\theta,\psi)$.
Note that the first of these differs from \gm.

\appendix{B}{Volume Contributions to Index Theorem}

{}For the general metric
\twnine\ on the Atiyah-Hitchin manifold $\mtilz$ we choose a vierbein
\eqn\theight{  e^{r}  = f(r) dr, \qquad
                                          e^{1}  = a(r) \sigma_1^R, \qquad
                                          e^{2}  = b(r) \sigma_2^R, \qquad
                                           e^{3}  = c(r) \sigma_3^R.  }
We will use $\mu=e^r\wedge e^1\wedge e^2\wedge e^3$ as
defining a positive orientation.
Note that
\eqn\ant{
\eqalign{\mu=&fabc \sin\theta d\theta \wedge d\phi\wedge  d\psi\wedge  dr\cr
=&{\sqrt g} d\theta \wedge d\phi\wedge  d\psi\wedge  dr.\cr}}
since $fabc>0$.
The spin connection is given by
\eqn\thnine{ \eqalign{ &\omega^{1 2}  = (1 + c'/f) \sigma_3^R,\qquad
                              \omega^{ 3 1}   =  (1+ b'/f) \sigma_2^R, \qquad
                              \omega^{2 3}  = (1+a'/f) \sigma_1^R, \cr
                             &\omega^{r 1} = -(a'/f) \sigma_1^R, \qquad
                               \omega^{r 2}  = -(b'/f) \sigma_2^R, \qquad
                              \omega^{r 3}  = -(c'/f) \sigma_3^R. \cr }}
where we have used
\eqn\bee{
{a'\over f}={(b-c)^2-a^2\over 2 bc } \qquad {\rm and}\quad{\rm cyclic}}
following from anti-self-duality and the fact that the $SO(3)$
action rotates the complex structures (see \ah).

The gravitational volume contribution to the index is
\eqn\ccb{ I_V^R =  {1 \over 192 \pi^2}
\int_\mtilz \tr R \wedge R = -{1 \over 48 \pi^2} \int_\mtilz
 \left( R^{r1} \wedge R^{r1} + R^{r2}
\wedge R^{r2} + R^{r3} \wedge R^{r3}  \right),}
where the anti-self-duality of the curvature
($R^{r1} = - R^{23} $ etc.) has been
used.  In terms of $\ahat = a'/f $, $\bhat = b'/f$, and $\chat = c'/f$ the
curvature components are
\eqn\ccc{\eqalign{ R^{r1} & = -\ahat' dr
\wedge \sigma_1^R + [ -\ahat + \bhat +\chat +
2 \bhat \chat ] \sigma_2^R \wedge \sigma_3^R \cr
R^{r2} & = -\bhat' dr \wedge \sigma_2^R +
[ \ahat - \bhat + \chat + 2 \ahat \chat ]
\sigma_3^R \wedge \sigma_1^R \cr
R^{r3} & = - \chat' dr \wedge \sigma_3^R +
[ \ahat + \bhat - \chat + 2 \ahat \bhat ]
\sigma_1^R \wedge \sigma_2^R \cr }}
which gives
\eqn\ccd{\eqalign{
I_V^R =&  + {1 \over 48 \pi^2} \int_\mtilz G(r) \sigma_1^R
\wedge \sigma_2^R \wedge
\sigma_3^R\wedge dr\cr
 =&  - {1 \over 48 \pi^2}\int \sin\theta d\theta d\phi d\psi
 \int_\pi^\infty G dr\cr}}
with
\eqn\cce{ G(r) = \left[ \ahat^2 + \bhat^2 +
\chat^2 - 2 \ahat \bhat - 2 \bhat \chat -2 \ahat \chat
- 4 \ahat \bhat \chat \right]'. }
Using the asymptotic forms for $\hat a$, $\hat b$, and $\hat c $
obtained form \asymp\ and \asympt\ we find
\eqn\ccf{
 I_V^R = {1 \over 24 \pi^2}
{1 \over 2} \int_0^\pi \sin \theta d
\theta \int_0^{2 \pi} d \phi \int_0^{2 \pi} d \psi =
{1\over 6}. }
The factor of $1/2$ arises from the $I_x$ identification.
Thus the gravitational volume contribution to the
Dirac index is $+1/6$.
As a check we can now calculate the Euler number:
\eqn\eul{
\chi=-{1\over 16\pi^2}\int_\mtilz \tr R\wedge *R=2,}
where we have used the anti-self-duality condition. This is consistent
with
the fact that $\mtilz$
contracts onto a two sphere.

{}For the volume gauge contribution, we have the field strength \senform\
\eqn\ccg{ F = d \alpha \wedge \sigma_1^R +
\alpha \sigma_2^R \wedge \sigma_3^R }
so that for a charge $q$ field we have
\eqn\cch{  I_v^F =   -{q^2 \over 8 \pi^2}
\int_\mtilz  F \wedge F = -{q^2 \over 8 \pi^2} \int_\mtilz
d(\alpha^2) \sigma_1^R \wedge \sigma_2^R \wedge \sigma_3^R =
 -{q^2 \over 8}, }
where we have used $\alpha(\pi) = 1/2$ and that $\alpha$ falls off
exponentially with $r$ as can be deduced from \difff.
Thus the sum of the volume terms for
a charge $q$ fermion is
\eqn\cci{ I_V^R + I_V^F = {1 \over 6} - {q^2 \over 8 }. }

\appendix{C}{$\eta$ invariants}

In this appendix we discuss the computation of the $\eta$
invariant, $\eta(0)$,  for the boundary of the
$O(2)$ bundle over the Atiyah-Hitchin manifold.
$\eta(0)$ is defined to be the analytic continuation
to $s=0$ of
\eqn\qq{ \eta(s) = \sum_{\lambda \ne 0}  \lambda^{-s} {\rm sign} \lambda .}
where the $\lambda$ are the eigenvalues of the Dirac operator on the
boundary of the manifold coupled to the flat $O(2)$ connection.
At infinity we have $a\approx b\approx r$ and
$c\approx -2$ and so the boundary metric is
\eqn\pp{ ds^2 =  r^2 ((\sigma_1^R)^2 + (\sigma_2^R)^2 ) + 4
(\sigma_3^R)^2 .}
{}For the moment we ignore the identification by $I_x$ and let $\psi$
run from $0$ to $4 \pi$ so that \pp\ defines a left-invariant (but not round)
metric on $S^3$. Moreover, we also ignore the flat $O(2)$ connection for
the moment.

$\eta(0)$ is invariant under conformal rescalings of the metric so in order to
compare with \ref\pope{C. N. Pope, J. Phys. A: Math. Gen. {\bf 14} (1981)
L133} we will consider the metric
\eqn\rr{ ds^2 = {1 \over 4} ((\sigma_1^R)^2 + (\sigma_2^R)^2) +
{\mu^2 \over 4} (\sigma_3^R)^2 }
in the limit $\mu \rightarrow 0$.
We then have the dreibein
\eqn\ss{e^1 = \hf \sigma_1^R, \quad e^2 =
\hf \sigma_2^R, \quad e^3 = {\mu \over 2} \sigma_3^R }
and spin connections
\eqn\tt{ \omega^{12} = (1-\mu^2/2) \sigma_3^R, \quad \omega^{31} = (\mu/2)
\sigma_2^R, \quad
\omega^{23} = (\mu/2) \sigma_1^R .}
The Dirac equation is
\eqn\uu{i \gamma^a {E^\mu}_a (\partial_\mu + {1\over
4}\omega_{\mu ab}\gamma^{ab}) \psi = \lambda \psi }
The $E_a = {E^\mu}_a \partial_\mu$ are dual to the $e^a$
and using  \ss\ we have
$E_1 = 2 \xi_1^R$, $ E_2 = 2 \xi_2^R$, $E_3 = 2 \xi_3^R /\mu$ where the
$\xi_i^R$ were introduced in \livf.
Substituting \am, the Dirac
equation reads
\eqn\xx{-2 \pmatrix{
\mu^{-1} L_3^R & L_-^R \cr
L_+^R & - \mu^{-1} L_3^R  \cr
} \psi  - {1 \over 2 \mu} (\mu^2 +2) \psi = \lambda \psi. }
The eigenfunction can now be constructed in terms of
the Wigner function $D^j_{m,s}(g)$
introduced in \wig.
Using \ralg, the eigenfunctions of the Dirac operator are of two types.
The first type is of
the form
\eqn\aaa{ \psi^{0 j}_{m,s} = \left (\matrix{
a D^j_{m,s} \cr
b D^j_{m,s+1} \cr
}\right ) }
with $s=-j, \ldots, j-1$. In order to be an eigenfunction we must have
\eqn\bbb{{b \over a} = -{2s+1 \over 2 \mu X}
\pm {1 \over 2 \mu X} \sqrt{(2s+1)^2 + 4 \mu^2 X^2} }
with $X^2 = j(j+1) - s(s+1)$.  Note that as a function of $s$ at fixed
$j$ we have
\eqn\bbba{ {b \over a}(s) = {a \over b} (-s-1) . }
The eigenvalues are
\eqn\ccc{ \lambda_\pm = -{\mu \over 2} \mp {1 \over \mu}
\sqrt{ (2s+1)^2 + 4 \mu^2 X^2 }. }
Since the $\eta$ invariant is left unchanged by a rescaling of all of
the eigenvalues,
to evaluate it in the limit $\mu \rightarrow 0$ we
rescale by a factor of $\mu $ and then
let $\mu \rightarrow 0$. In this limit we get for the rescaled eigenvalues
\eqn\ddd{ \lambda_\pm = \mp (2s+1) . }
Since these are symmetric between positive and negative eigenvalues they
do not contribute to $\eta(s)$ in the limit $\mu \rightarrow 0$.

The second type of eigenfunctions have  the form
\eqn\eee{
\psi^{+j}_m = \left (\matrix{
D^j_{m,j} \cr
0\cr } \right ) }
and
\eqn\eeea{\psi^{-j}_m = \left (\matrix{
0\cr
D^j_{m,-j} \cr }\right ) }
with eigenvalues
\eqn\fff{ \lambda_+ = \lambda_-  = -{\mu \over 2}  - {1 \over \mu} (2j+1). }
After the rescaling by $\mu$ and taking the limit we get eigenvalues
$\lambda_\pm = - (2j+1)$. These eigenvalues are all negative.
So we get eigenvalues $-(2j+1)$
with degeneracy $2(2j+1)$ with the $2$ from $\psi_\pm$ and the $(2j+1)$
from the possible $m$ values. The $\eta$ invariant is thus
\eqn\ggg{ \eta(s) = - \sum_j 2 (2j+1) (2j+1)^{-s} }
and since $j$ takes on half-integer and integer values this is equal to
\eqn\hhh{ \eta(s) = -2 \sum_{p=1}^\infty p^{-s+1} \equiv -2 \zeta (s-1) }
and $\eta(0) = -2 \zeta(-1) = 1/6$.
Note that in the limit $\mu\to 0$ there are no harmonic spinors and
hence $h_D=0$.

So far we have just redone the standard computation of the
$\eta$ invariant on the squashed
$S^3$ \hitchin\ in the limiting case $\mu\to 0$.
 In extending this computation to
the boundary of the Atiyah-Hitchin manifold there are two complications
which we must deal with. The first is that the boundary is not $S^3$
but rather $SO(3)/Z_2 \equiv S^3/Z_4$ because of the identification
by $I_x$. As a result the boundary has different inequivalent spin structures.
Since the Atiyah-Hitchin manifold is simply connected and
consequently has a unique spin structure only one
of these spin structures can extend in smoothly to the interior. Determining
the spin structure at the boundary appears to be somewhat subtle, luckily
we will be able to determine its effect by consistency requirements. The second
complication arises from the $O(2)$ connection. The field strength
of this connection falls off exponentially leaving a flat $O(2)$ connection
at the boundary.  The holonomy of this connection at the boundary
has been computed in $\ms$ with the result that charge $q$ fermions
pick up a phase of $e^{\pm i q \pi /2 }$.

Now let us first consider spinors with $q=0$.
Identifying under the action of $I_x$ leads to $\pi_1$ of the
boundary being $Z_4$.
The different spin structures on the boundary can be specified
by giving the phase picked up by the fermion after propagating
between two points related by $I_x$.
Since $I_x$ is a right action it commutes with the isometries
$SO(3)_L $ and consequently the phases picked
up by the fermions fields
cannot depend on $m$.
As in the previous calculation the eigenfunctions \aaa\ have no spectral
asymmetry in the limit $\mu\to 0$ and thus do not contribute to $\eta(0)$.
We  now consider the eigenfunctions \eee\ and \eeea.
{}For a given $j$ a single eigenfunction
picking up a phase  $e^{2\pi i r \over N}$ (in our case
$N=4$ since $\pi_1$ of the boundary is $Z_4$ ) contributes
a term $ -f(r,N,-1) $ to the $\eta$ invariant where we have
introduced the function
\eqn\one{ f(r,N,s)= \sum_{n= r ~{\rm mod} ~N}^\infty n^{-s} }
with $r$ and $N$ integers.
We have
\eqn\two{ f(r,N,s) = \sum_{n=0}^\infty  (n N + r )^{-s}
= N^{-s} \sum_{n=0}^\infty (n+r/N)^{-s}
= N^{-s} \zeta(s,r/N) }
in terms of  the generalized
Riemann zeta function.  The value we need may be evaluated with
the result \gradry
\eqn\three{
\eqalign{  f(r,N,-1) &= N \zeta(-1,r/N) = - N {  B'_{3} (r/N) \over 6}
\cr &=
-N\left( {1 \over 12} +
{1 \over 2} ((r/N)^2-(r/N))\right) \cr}}
(it is amusing to note that the same sum arises in the computation of the
dimension of twist fields for orbifold compactifications of string theory
\dhvw).
Assuming in general that the two eigenfunctions
\eee\ and \eeea\ pick up phases $r_1$ and $r_2$
due to the spin structure
at infinity,
they will give a contribution to the $\eta$
invariant of $-f(r_1, 4,-1) - f(r_2,4, -1)$.
Noting the values $f(0,4,-1) = -1/3$,
$f(1,4,-1) = f(3,4,-1) =1/24$ and $f(3,4,-1) = 1/6 $
we see that the only possibility which
is consistent with an integer index and the vanishing theorem for $q=0$
mentioned in the text is $r_1=r_2 = 2$.

Given this indirect evaluation of the effect of the spin structure for $q=0$
it is straightforward to generalize the calculation to non-zero $q$. Using
the fact that the holonomy is $e^{\pm \pi i q /2}$  there are two
possibilities: either the two eigenfunctions pick up the same phase or
opposite phases. The first possibility leads to
the $\eta$ invariant: $- 2f([q+2]_4 ,4,-1)$, where $[q+2]_4$
is $(q+2)$ mod 4 (since the phase is $\pi i(q+2)/2$).
The second possibility
leads to the $\eta$ invariant: $-f(q+2,4,-1)-f(2-q,4,-1)$
but these two expressions are equal. We thus
conclude that the $\eta$ invariant is given by
\eqn\here{
\eta_{q} (0)=
 {2 \over 3} + {1 \over 4}( [q+2]_4^2 - 4[q+2]_4).}
Noting that $h_D$ still vanishes,
this then leads to the results stated in the text.

\listrefs
\end